\documentclass{article}
\usepackage{epsf}
\usepackage{emulateapj}
\input{psfig}
\usepackage{mathptm}

\def\sgra	{Sgr~A$^*$}
\def\msol   {{M$_{\odot}$}}
\def\mo     {{M$_{\odot}$}}
\def\kms    {~km~s$^{-1}$}
\def\gax    {${_>\atop^{\sim}}$}
\def\lax    {${_<\atop^{\sim}}$}

\def\etal   {{\sl et~al.}}
\def\ergs   {~erg~s$^{-1}$}
\def\cs     {~{c~s$^{-1}$}}
\def\ergscm2s  {~erg~cm$^{-2}$~s$^{-1}$}
\def\cm2s   {~cm$^{-2}$~s$^{-1}$}
\def\cm2   {~cm$^{-2}$}

\def\mathfont#1{\ifmmode{#1}\else{$#1$}\fi} 
\def\lae{\mathrel{<\kern-1.0em\lower0.9ex\hbox{$\sim$}}}  
\def\gae{\mathrel{>\kern-1.0em\lower0.9ex\hbox{$\sim$}}}  
\def\kms{\ifmmode{{\rm km\ s}^{-1}}\else{${\rm km\ s}^{-1}$}\fi}

\def\msun{\ifmmode{\ {\rm M}_\odot}\else{$ {\rm M}_\odot$}\fi}  
\def\msunyr{\ifmmode{\msun \ {\rm yr}^{-1}}\else{$\msun \ {\rm yr}^{-1}$}\fi}

\def\ref#1{\noindent\hangindent=24.0pt\hangafter=1{#1}\par}
\def\la{\hbox{\rlap{$<$}\lower.5ex\hbox{$\sim$}\ }}
\def\ga{\hbox{\rlap{$>$}\lower.5ex\hbox{$\sim$}\ }}

\lefthead{Garcia et al.}
\righthead{A First Look at M31 with Chandra}

\slugcomment{to be submitted to {\em The Astrophysical Journal Letters}}

\begin{document}

\title{A First Look at the Nuclear Region of M31 with Chandra}

\author{Michael R. Garcia\altaffilmark{1}, 
Stephen S. Murray \altaffilmark{1}, 
Francis A. Primini \altaffilmark{1}, 
William R. Forman \altaffilmark{1}, 
Jeffrey E. McClintock \altaffilmark{1}, 
and Christine Jones \altaffilmark{1} }

\altaffiltext{1}{Harvard-Smithsonian Center for Astrophysics, MS-4, 60
Garden St., Cambridge, MA 02138; email: garcia,ssm,fap,wrf,jem,cjf,@head-cfa.harvard.edu}

\begin{abstract} 

We report on the first observation of the nuclear region of M31 with
the Chandra X-ray Observatory.  The nuclear source seen with the
Einstein and ROSAT HRIs is resolved into five point sources.  One of
these sources is within $1''$ of the M31 central super-massive black
hole.  As compared to the other point sources in M31, this nuclear
source has an unusual x-ray spectrum. Based on the spatial coincidence
we identify this source with the central black hole, and note that the
unusual spectrum is a challenge to current theories.  A bright
transient is detected $\sim 26''$ to the west of the nucleus, which
may be associated with a stellar mass black hole.

\end{abstract}

{\it Subject headings:}  Galaxies: individual (M31) -- black holes 

\vfill

\section{Introduction}

As our nearest Milky Way analog, M31 offers us a chance to study a
galaxy like our own without the obscuring effects of living in the
middle of the Galactic plane.  For example, the nucleus of our Galaxy
(\sgra), is obscured by $\sim 30$ magnitudes of visual extinction
(Morris and Serabyn 1996), while the nucleus of M31 likely suffers
\lax~2 magnitudes of extinction (see Section 2.3.1).  In addition, the study
of x-ray binaries in the Galactic plane is hindered by reddening 
sometimes reaching $>10$~magnitudes, which can be compared to an average
${\rm E(B-V) = 0.22}$~magnitudes for globular clusters in M31 (Barmby \etal\/
2000).

Ground based measurements of the rotational velocity of stars near the
core of M31 provide strong evidence of a central dark, compact
object of mass $3.0 \times 10^7$\msol, presumably a black hole
(Kormendy and Bender 1999 and refs therein).  
HST observations resolved the M31 nucleus
into two  components (P1 and P2)
separated by $\sim 0.5''$ (Lauer \etal\/ 1993).
These
observations support the model of the double nucleus of M31 as a torus
of stars orbiting the core in a slightly eccentric orbit (Tremaine
1995).  Post COSTAR HST observations have shown that there is a group of
partially resolved UV-bright stars between P1 and P2 at the
position of the central black hole (Brown \etal\/ 1998).

The first identification of an x-ray source with the M31 nucleus came
with Einstein observations, which found a source within $2.1''$ of the
nucleus with ${\rm L_x = 9.6 \times 10^{37}}$\ergs (0.2-4.0~keV, Van Speybroeck
\etal\/ 1979).  While this source was not variable in this first
observation, subsequent Einstein observations showed the nucleus to be
variable by factors of $\sim 10$ (Trinchieri and Fabbiano 1991) on
timescales of 6~months.  Published ROSAT observations show ${\rm L_x} =
2.1 \times 10^{37}$\ergs, which is at the faint end of the Einstein range
(Primini, Forman and Jones 1993).

Radio observations reveal a weak ($\sim 30 \mu~{\rm Jy}$) source at the
core (Crane, Dickel and Cowan 1992). The luminosity at 3.6~cm is $\sim
1/5$ that of \sgra, a puzzle given that the M31 nucleus is $\sim
30$ times more massive (Melia 1992).  The correlation between the radio and
x-ray properties of low-luminosity super-massive black holes (Yi
and Boughn 1999) might be explained by an ADAF
model, but M31 is an outlier in these correlations. 

The point sources distributed throughout M31 are likely x-ray binaries
and supernova remnants similar to those in our galaxy.  The fact that
$\sim 40$\% of these sources are variable is consistent with this
hypothesis (Primini, Forman and Jones 1993).  As in the galaxy, some
of these point sources are transient.  Comparison of Einstein and
ROSAT images shows that $\sim 6$\% of the sources are transient
(Primini, Forman and Jones 1993). A comparison of Einstein and EXOSAT
observations allowed discovery of two transients (White \& Peacock 1988), and a
study of the ROSAT archive allowed discovery of a supersoft x-ray
transient (White \etal\/ 1995).

The sensitivity and high spatial resolution of Chandra (van Speybroeck
\etal 1997, Weisskopf and O'Dell 1997) provide new insights into the
x-ray properties of M31.  A few of those new insights, concerning the
nucleus and a new transient, are reported in this letter.

\section{Observations}

\subsection{Chandra}

Chandra was pointed at the nucleus of M31 for 17.5~ks on Oct 13, 1999.
This pointing occurred immediately before Chandra operations paused for
the passage through the Earth radiation belts, and the radiation
environment was already higher than average.  This caused high counting
rates in the ACIS-S3 chip, which saturated telemetry and caused data
dropouts.  The S3 counting rate was used as an indicator of high
background, and whenever it increased beyond 1.5\cs we rejected
the data.  Consequently we obtained 8.8 ks of active observing time.

The standard four ACIS-I (Garmire \etal 1992) chips were on; therefore
a \hbox{$\sim 16' \times 16'$} region of the center of M31 was
covered.  In this letter we concentrate on the observations of the
central $\sim 1'$ only.  The image of this nuclear region is shown in
Figure~1.

Data were analyzed with a combination of the CXC Caio V1.1 (Elvis
\etal 2000), HEASARC XSPEC V10.0 (Arnaud 1996), and software written
by Alexey Vikhlinin (Vikhlinin \etal\/ 1998).  Unless otherwise
specified, all error regions herein are 68\% confidence bounds and
include a 20\% uncertainty in the ACIS effective area below 0.27~keV.
We note that this calibration uncertainty is $<50$\% of the
statistical uncertainties for the sources considered herein.

\subsection{ROSAT}

ROSAT imaged the central region of M31 six times from 1990 to 1996, with
exposure times ranging from 5~ks to 84.5~ks (see Primini, Forman and
Jones 1993. and Primini \etal\/ 2000).  The last of these exposures
was 84.5~ks in 1996 January.  The image of the nuclear region from
this observation is shown in Figure 1 (top).


\begin{center}

\psfig{file=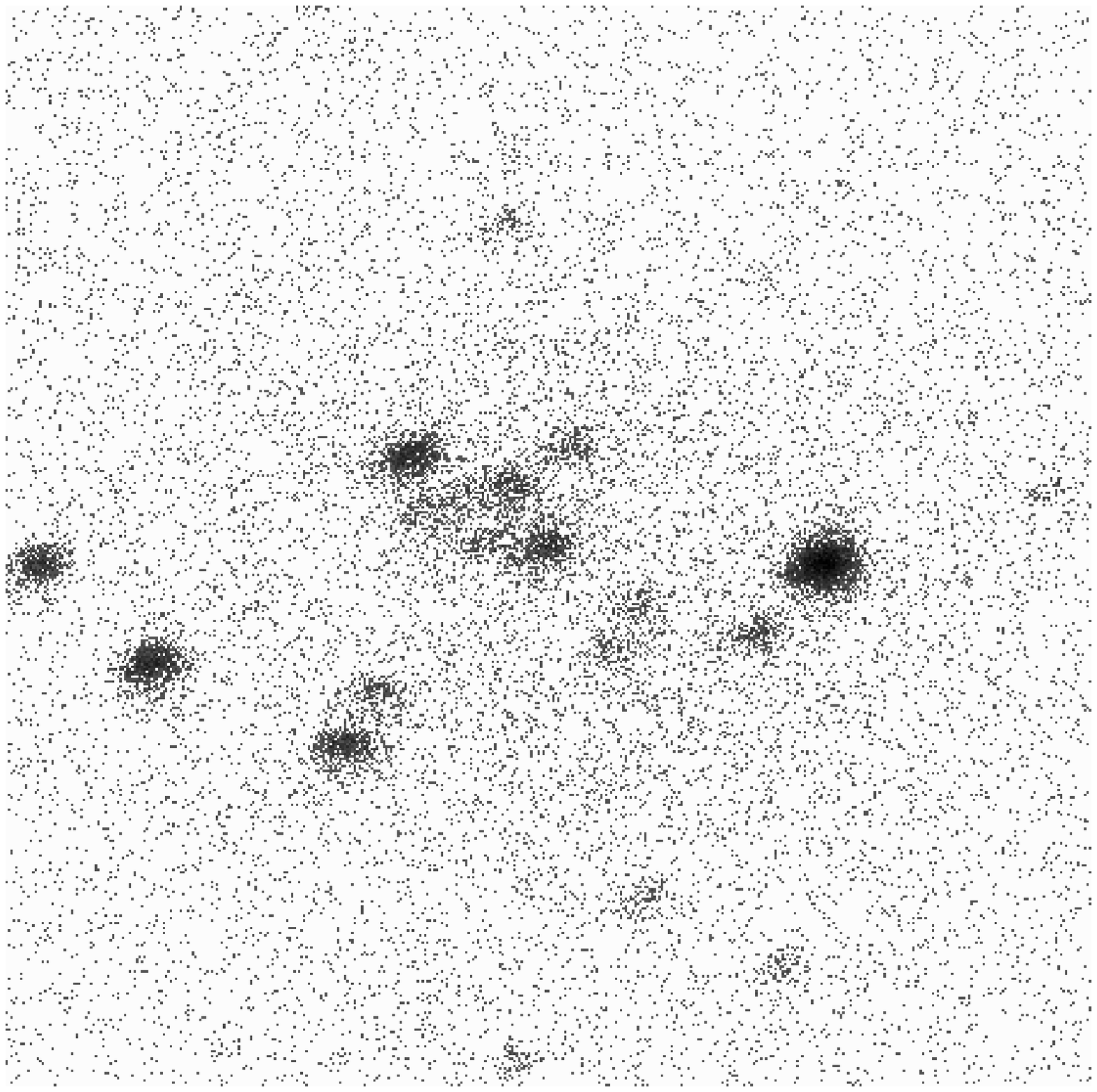,width=3in}
\psfig{file=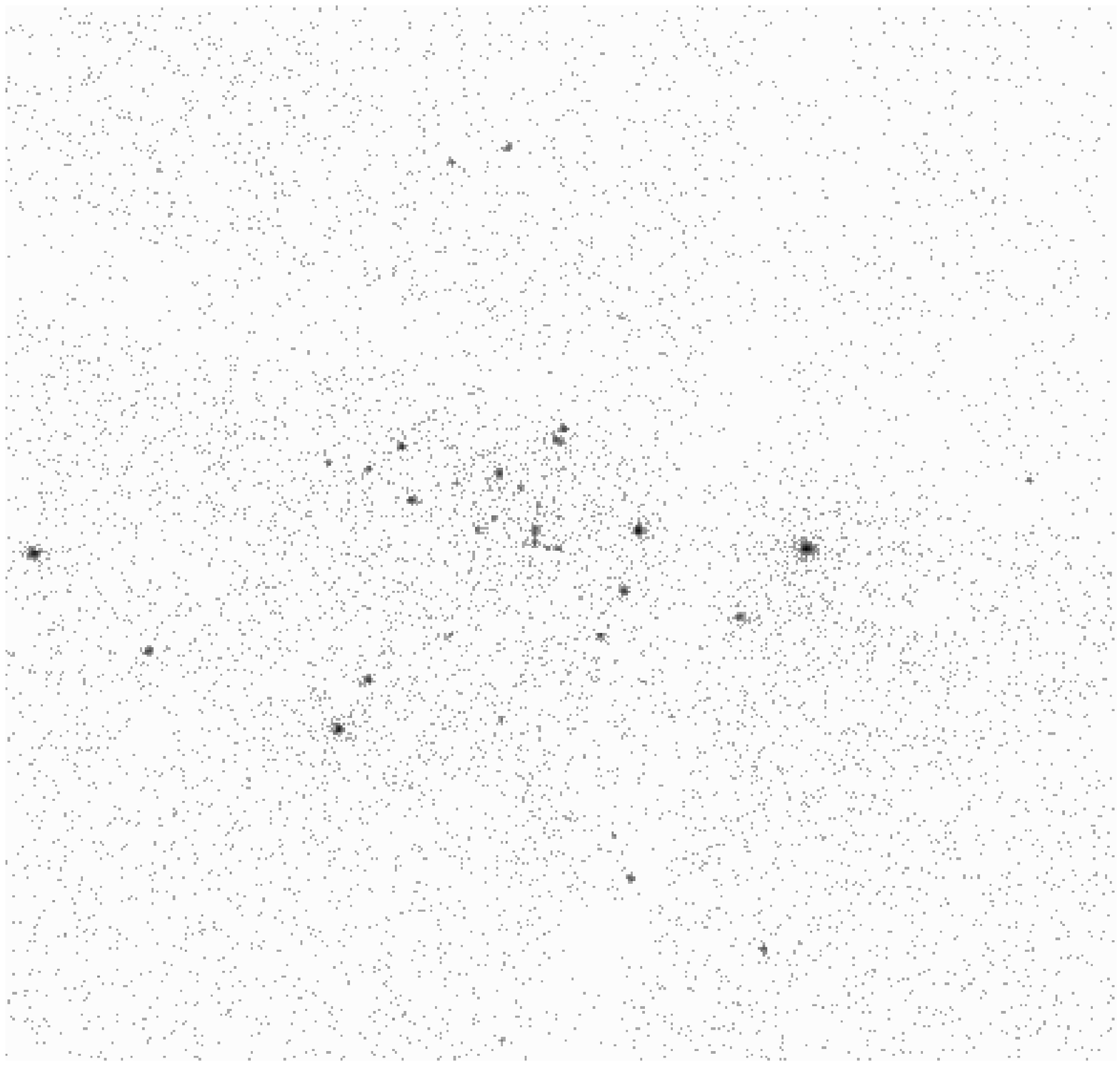,width=3in}
\begin{minipage}[h]{3.5in}
{\small
Figure 1:  Top:  The nuclear region of M31 as it appears in an 84.5~ks
ROSAT HRI observation in January 1996. Bottom:  The same as seen in an 8.8~ks 
Chandra ACIS-I observation on Oct 13, 1999.  The cross-like shadow
seen in the ACIS-I observation is due to the gaps between the 4 ACIS-I
chips.  These images are 4~arcmin on a side.}

\end{minipage}

\end{center}

\subsection{Data Analysis}

 
{\bf The Nucleus:}~ The central object seen with the ROSAT HRI is clearly resolved into 5
sources (Figure 2).  The Chandra aspect solution is based on 5~stars
from the Tycho (Hipparcos) catalog, so has the potential to be good to
a few tenths of an arc-sec (Aldcroft \etal 2000).  Based on the aspect
solution alone, we find that one of these five sources, CXO
J004244.2+411608, is within $< 1''$ of the position of the radio
nucleus (Crane \etal\/ 1992).  As an independent check on the aspect,
we computed a plate solution for the x-ray image using the positions
of 10 x-ray detected globular clusters from the Bologna catalog
(Battistini \etal\/ 1987).  This solution has an uncertainty of $\sim
0.7"$ rms in RA and Dec, and agrees (within the errors) with the
Chandra aspect.

\begin{center}

\psfig{figure=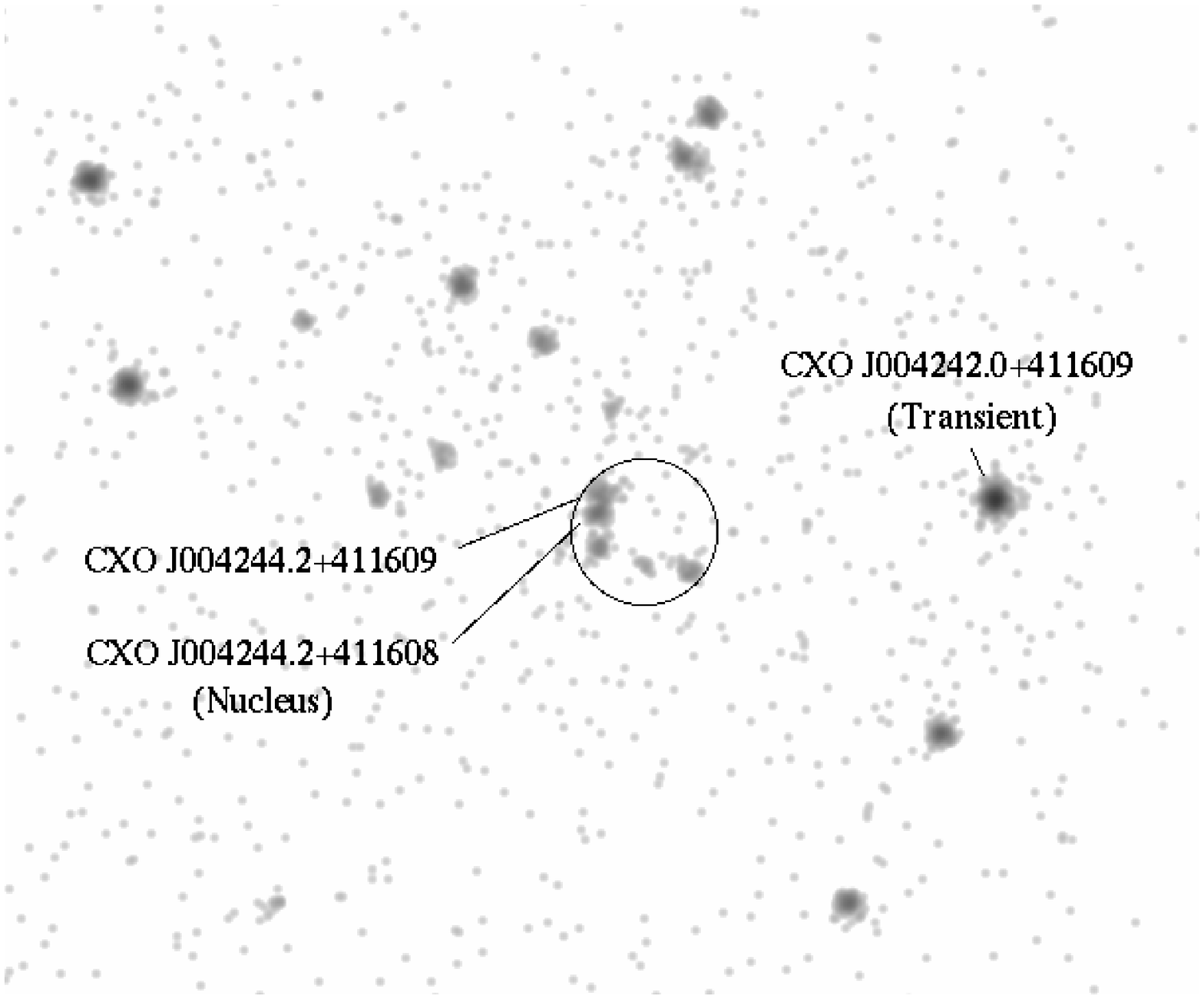,width=3in,angle=-90,width=3in}
\begin{minipage}[h]{3.5in}
{\small
Figure 2: An enlargement of Figure 1(bottom), showing the nuclear region in
detail.  The circle surrounding the central sources is $5''$ in
diameter, approximating the resolution of the ROSAT HRI.  This image
is 1~arcmin on a side. 
}
\end{minipage}

\end{center}

In order to get a first look at the spectra of the point sources, we
performed a wavelet deconvolution (Vikhlinin \etal\/ 1998) of the image and
found  121~point sources in the full $16' \times 16'$ FOV of ACIS-I
(these sources will be discussed in a separate paper).  We then
computed the hardness ratio of the 79 sources with more than
20~counts.  In the histogram of this ratio (Figure~3) 
the nuclear source is one of three outliers with extremely soft
spectra.  The fact that the nuclear spectrum is distinctly different
from the mean may indicate that there is something fundamentally 
different about this source.  

\begin{center}

\psfig{figure=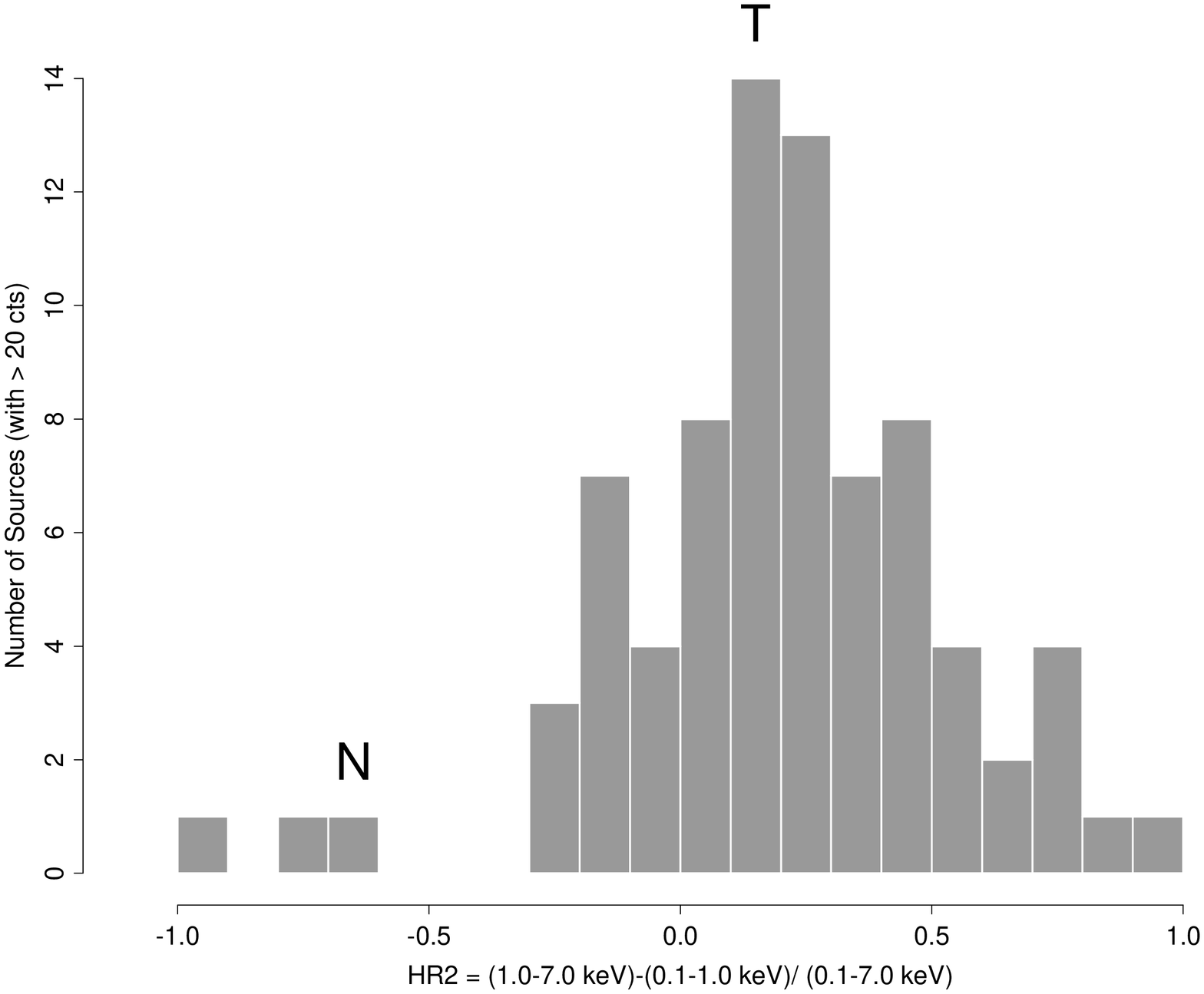,width=2.5in,angle=-90}
\begin{minipage}[h]{3.5in}
{\small
Figure 3: The hardness ratio for 79 sources with $>20$ total counts
found in the ACIS-S image of M31. The nuclear source,
CXO~J004244.2+411608, has the third lowest hardness ratio, and is
indicated by the ``N''. The nearby transient is indicated by the
``T''. The source $1''$ North of the nucleus is in the first bin below
0.0, the source $\sim 1.5''$ to the South of the nucleus is in the bin
indicated by the ``T''.}
\end{minipage}

\end{center}

We extracted 100~counts from a $3$~square-arc-sec region surrounding
the nucleus.  In order to limit contamination from
CXO~J004244.2+411609, which is only $1.0"$ to the North, we excluded
photons more than $0.5"$ to the North of the nuclear source. 
The resulting PHA spectrum was fit with XSPEC, after first
binning the data such that each fitted bin had $>10$ counts.  Gehrels
weighting was used for the fits (Gehrels 1986).  The fits were
limited to the 0.2-1.5~keV region, as there were insufficient counts
outside of this region.

Simple models (powerlaw, black-body, bremsstrahlung, with interstellar
absorption) provide acceptably good fits to the data.  The power law
fits find a slope $\alpha = 5_{-2.4}^{+7}$, and limit ${\rm N_H =
4_{-3.5}^{+9} \times 10^{21}}$\cm2 .  In order to reduce the error
range on the fitted slope we choose to limit the allowed range of
absorption to that found for the nearby transient (below), ie, to
${\rm N_H = 2.8\pm 1.0 \times 10^{21}}$\cm2 .  This then allows us to
further restrict the slope (or temperature) of the spectrum to $\alpha
= 4.5 \pm 1.5$, kT$= 0.15^{+0.06}_{-0.03}$, or kT$= 0.43 \pm 0.17$ for
power-law, black-body or bremsstrahlung fits (respectively).

The detected 0.3-7.0~keV flux, assuming the further restricted range of
parameters for the power law model, is $5.8^{+0.9}_{-0.5} \times
10^{-14}$ \ergscm2s , corresponding to an observed luminosity of
$3.9^{+0.6}_{-0.3} \times 10^{36}$\ergs at 770~kpc (Stanek and
Garnavich 1998).  At the lowest ${\rm N_H}$ and flattest $\alpha$ in
this range, approximately 60\% of the 0.3-7.0 keV flux is absorbed by
the ISM, while at the highest ${\rm N_H}$ and steepest $\alpha$,  nearly
98\% of the flux is absorbed.  The corresponding emitted luminosity
ranges from $1.2 \times 10^{37}$\ergs\/ to $1.6 \times 10^{38}$\ergs,
and has a nominal value at the best fit parameters of $4.0 \times
10^{37}$\ergs .

In order to test our assumption that the ${\rm N_H}$ measured for the
transient is appropriate to apply to the nucleus, we fit power law
spectra to four other bright nearby sources.  These sources are all
further away from the nucleus, with distances ranging from $30''$ to
$2'$, and have between 237 and 823 detected counts.  In every case the
90\% confidence regions for ${\rm N_H}$ overlap with the transient.
Given that there is no evidence for large variations in ${\rm N_H}$ in
the region around the nucleus, it is reasonable to assume the nuclear
${\rm N_H}$ is the same as that of the transient.  Note that the
galactic ${\rm N_H} \sim 7 \times 10^{20}$\cm2 ~in the direction of
M31 (Dickey \& Lockman 1990), so our results are consistent with
additional local absorption within M31 itself.  If the gas/dust ratio
in M31 is similar to that in the Galaxy, the nuclear ${\rm A_V = 1.5
\pm 0.6}$ (Predehl and Schmitt 1995).


{\bf The Nearby Transient:}~ We extracted 763~counts for a $1"$ radius circle at the position of
CXO~J004242.0+411608.  This data was similarly grouped into bins with
$>10$ counts, and fit to simple models with XSPEC. Chi-squared fitting
with Gehrels weighting was used to find the minimum chi-squared
spectral parameters.  Power law, bremsstrahlung, 
and blackbody fits are all acceptable  
($\chi^2/\nu < 1.13$ for 71 DOF), but the power law fits produce the 
lowest $\chi^2/\nu  \sim 0.56$. 
Significant counts are seen out to 7.0~keV.  The best fitting power
law number slope is $1.5 \pm 0.3$, with a best fit ${\rm N_H = 2.8 \pm
1.0 \times 10^{21}}$\cm2 .  

Bremsstrahlung and black body fits formally allow ${\rm N_H = 0}$\cm2 , but
as the Galactic value to M31 is ${\rm N_H = 7 \times
10^{20}}$\cm2 , we restrict the fitting space to values larger
than this.  Bremsstrahlung fits are not able to set an upper limit to
the temperature, but set a lower limit of kT$>6$~keV.  Black body fits
limit the temperature to kT$=0.75 \pm 0.25$~keV.  Assuming a power law
model, the detected flux is $ 7.4 \pm 0.7 \times 10^{-13}$\ergscm2s ,
corresponding to a observed luminosity of $5.1 \pm 0.5 \times
10^{37}$\ergs, and an emitted luminosity of $7.0 \pm 0.8 \times
10^{37}$\ergs (0.3-7.0~keV).  The hardness ratio is typical of
other point sources (Figure~3).

We examined each of the 5 ROSAT HRI observations of the center of M31,
and find that there is no source apparent at the position of this
transient in any of these exposures.  For the deepest (and last)
observation, we find 78 counts in a a $7.5''$ arcsec radius at the
position of this transient, which is consistent with the background
caused by the diffuse emission in M31 (Primini \etal\/ 1993).  From
this we compute a 95\% (2 $\sigma$) upper limit of 17.7~counts.
Assuming the power law spectrum determined above for this source in
outburst, and applying a small correction for the flux not contained
in the $7.5''$ circle, this corresponds to an upper limit to the
emitted luminosity of the source of $3.0 \times 10^{36}$\ergs in the
0.3-7.0~keV band.  Thus the transient brightened by at least a factor
of $\sim 20$.

\subsection{Discussion}


{\bf The Nucleus:}~ Several authors have previously noted the unusual x-ray and radio
luminosity of the nucleus of M31 (Melia 1992, Yi and Boughn 1999). 
We note that the x-ray luminosity we find herein is substantially
lower than that quoted in several recent papers comparing x-ray and
radio luminosities of low luminosity super-massive black holes
(eg, Franceschini, Vercellone and Fabian 1998, Yi and Boughn 1999).
At this revised luminosity the M31 nucleus appears to be even more of 
an outlier on the correlations between radio luminosity, x-ray
luminosity, and black hole mass found for low luminosity super-massive
black holes (Yi and Boughn 1999, Figures 4 \& 5). 

The unusual x-ray and radio luminosity has
lead to the suggestion that perhaps the source may not be 
associated with the central black hole, but is merely a chance
co-incidence (van Speybroeck \etal\/ 1979, Yi and Boughn 1999).  The
probability of a chance co-incidence depends upon what search region
one uses, and a posteriori, it is hard to know what the relevant
search region is.  If we use the full ACIS FOV as the search region,
then the chance of any one of the 121 detected sources source being
within $1''$ of the nucleus is $\sim 4 \times 10^{-4}$.  However, the
surface density of sources increases towards the nucleus, so the
chance probability may be higher than this.  If one limits the search
region to the $\sim 25$ square arc-sec area which contains the five
sources ROSAT and Einstein were not able to resolve, the the chance
probability is $\sim 20$\%.  This is most likely an overestimate, as
can been seen by carrying this argument to its extreme (and
non-sensible) limit:  if one limits the search region to the 1~square
arc-sec region around the nucleus, the chance that the one source
within that region is within $1''$ of the nucleus is 100\%!. 

While it may be unclear what the appropriate search region is, it
seems clear that a chance alignment cannot be dismissed out of hand.
This motivated us to search for other unusual characteristics of the
central source, which led to the discovery that it has an unusually soft
spectrum. We speculate that the unusual spectrum is due, at least in
part, to the high mass of the nucleus, and that the unusual spectrum may
provide a clue to the origin of the unusually weak radio emission. 

However, because there are no observational precedents or strong
theoretical arguments which would lead us to expect the spectrum of
the a $\sim 10^7$\mo low luminosity black hole
to be very soft, we cannot identify the unusual
spectrum as a signature of the central black hole.  Our
identification is based solely on the positional co-incidence, and the
unusual spectrum is left as a challenge to models.  

While previous Einstein and ROSAT observations are unable to separate
the nuclear source from the surrounding four sources, the fluxes
indicate that the nucleus (or surrounding emission) is highly
variable.  In order to compare these fluxes to the Chandra flux, we
assume the nuclear power law spectrum found above, and use the
counting rates from the literature (Van Speybroeck \etal\/ 1979,
Trinchieri \& Fabbiano 1991, Primini \etal\/ 1993) to calculate
0.2-4.0~keV detected fluxes.  The uncertainty in the nuclear spectrum
allows up to 40\% uncertainty in the conversion from counting rate to
flux.  In order to make a fair comparison, Table~1 lists 
the summed flux from all 5 nuclear sources in the Chandra image.

\vskip 8pt

\centerline{Table 1: M31 Nuclear X-ray Flux}
\centerline{
\begin{tabular}{|c|c|c|}\hline
Date	&	Observatory	&Flux~($10^{-13}$\ergscm2s ) \\\hline
1979 Jan	&Einstein	& $7.07\pm 0.06$	\\\hline
1979 Aug	&Einstein	& $0.60\pm 0.18$	\\\hline
1980 Jan	&Einstein	& $3.50\pm 0.64$	\\\hline
1990 July	&ROSAT		& $1.70\pm 0.12$	\\\hline
1999 Oct	&Chandra (5)	& $1.43\pm 0.15$	\\\hline
\end{tabular}
}
\vskip 8pt

Strong variability of unresolved sources is often cited as evidence
for a small number of sources, simply because it is more likely that a
single source varies rather than a group of sources varies
coherently. 
If we apply this argument to the M31 nucleus, it implies that one of
these five sources (perhaps the nucleus itself?) is highly variable.
It would then be appropriate to assume that the average flux of the
surrounding four sources is $\sim$constant, and subtract this flux
from the Einstein and ROSAT measurements in order to determine the
flux of the nucleus alone.  From the Chandra image, the flux from
these four sources is $0.85 \times 10^{-13}$\ergscm2s .  Subtracting
this, we see that the lowest Einstein flux measurement is consistent
with zero flux from the nucleus, and indicates a factor of \gax~40
variability. 

As an aside, we note that the detection of \sgra with Chandra
(Garmire 1999) does not necessarily rule out an M31-like spectrum.
The much higher ${\rm A_V \sim 30}$ to \sgra would reduce the observed
count rate from an M31-like spectrum by $\sim 60$ times, but the $\sim
100$ times smaller distance would more than make up for this.

Standard ADAF models are not able to explain the ratio of x-ray to
radio luminosity of the nucleus (Yi and Boughn 1999).  However, models
including winds (Di Matteo \etal\/ 1999) and/or convective flows
(Narayan, Igumenshchev \& Abramowicz 1999) may be able to explain this
ratio.  These models generally predict hard spectra in the x-ray
region, so may not be able to explain the extremely soft spectrum
reported herein (Quataert 2000, pc).  We note that the x-ray
luminosity of M31 is several orders of magnitude below that 
typically considered in these models, implying that the models may not
fully describe this parameter space. 


{\bf The Nearby Transient:}~ The nature of the bright transient is uncertain.  
By analogy to Milky Way sources, its transient nature and
luminosity imply that it is either a massive X-ray binary, typically
consisting of a Be-star and a pulsar, or an x-ray nova, often
consisting of a late-type dwarf and a black hole (White, Nagase and
Parmar 1995, Tanaka and Lewin 1995).  The spectral slope of $\alpha =
1.5$ is between the hard spectra typically seen in x-ray pulsars 
($ 0.0 < \alpha < 1.0$, White, Nagase and Parmar 1995) and the softer
spectra seen in x-ray novae in outburst ($\alpha \sim 2.5$, Ebisawa
\etal\/ 1994; Sobczak \etal\/ 1999).  At late times in the decay of an
x-ray novae the spectrum often hardens to $\alpha \sim 1.5$, but this
would imply that the peak outburst luminosity of this transient 
was \gax~$10^{39}$\ergs . 

The absorption of ${\rm N_H} = 2.8 \pm 1.0 \times 10^{21}$\cm2 is more
typical of x-ray novae than Be-star pulsar systems, which often have
${\rm N_H} > 10^{22}$.  Perhaps the strongest argument in favor of an
x-ray nova hypothesis is the location of the transient:  stars in the inner
bulge of M31 are likely old, disk/bulge population stars typical of
those in x-ray novae, rather than the young, Be stars typically found
in star forming regions and in Be-star pulsar systems.  

We note that in either case the optical magnitude of the transient in
outburst is likely to be V$\sim 22$, making the object visible with
HST.  An x-ray nova would be expected to show a large variation in V
from quiescence to outburst, while a Be-star pulsar would show a more
moderate variation.  HST observations are underway in an attempt to
clarify the nature of this transient.

\acknowledgments

We thank Pauline Barmby for providing results on M31 globular cluster
reddenings and positions prior to publication, Eliot Quataert for
comments on ADAF models, and the CXC team for help with ACIS data
reductions.  This work was supported in part by NASA Contract NAS8-39073.

\

\end{document}